\documentclass[review,12pt,authoryear]{elsarticle}

\usepackage[utf8]{inputenc}
\DeclareUnicodeCharacter{00A0}{ }
\usepackage{amssymb}
\usepackage{amsmath}

\usepackage[hidelinks]{hyperref}
\usepackage{multirow}

\usepackage{subcaption}
\usepackage{overpic} 
\usepackage{xcolor}
\usepackage{soul}
\usepackage{tabularx}
\newcolumntype{Y}{>{\raggedright\arraybackslash}X}
\newcolumntype{L}[1]{>{\raggedright\arraybackslash}p{#1}}

\usepackage{booktabs}
\usepackage{array}

\journal{Computers and Geotechnics}

\begin{document}

\begin{frontmatter}

\title{A Gaussian process coarse-grained potential for Na-montmorillonite}

\author[label1]{Yalda Pedram}
\author[label1]{Yaoting Zhang}
\author[label2]{Laurent Brochard}
\author[label3]{Chang Seok Kim}
\author[label1]{Laurent Karim Béland\corref{cor1}}
\ead{laurent.beland@queensu.ca}
\cortext[cor1]{Corresponding author.}


\affiliation[label1]{organization={Department of Mechanical and Materials Engineering, Queen’s University},
            city={Kingston},
            postcode={K7L 3N6},
            state={Ontario},
            country={Canada}}

\affiliation[label2]{organization={Laboratoire Navier, CNRS, Université Gustave Eiffel, ENPC, Institut Polytechnique de Paris},
            city={Marne-la-Vallée},
            country={France}}
            
\affiliation[label3]{organization={Nuclear Waste Management Organization},
            addressline={30 Adelaide Street East, Suite 600},
            city={Toronto},
            postcode={M5C 3H1},
            state={Ontario},
            country={Canada}}
\begin{abstract}

Hydraulic transport in compacted bentonite is diffusion-controlled and governed by the hydration and microstructure of sodium montmorillonite (Na--MMT). Experiments cannot resolve how platelet interactions govern pore structure, transport and stiffness, while existing coarse-grained models smooth hydration oscillations or require manual corrections. We develop a tabulated potential combining Morse interactions between platelet centre and edge sites with a Gaussian process regression correction trained on all-atom potentials of mean force. It captures the hydration-induced complexity of the potential-of-mean-force profiles, including the three-water (3--W) hydration minimum and transfers across geometries, layer-charge variants and unseen configurations. Applied to monodisperse and polydisperse Na--MMT assemblies at dry densities of 0.8--1.3~g\,cm$^{-3}$, the model captures the 3--W to 1--W transition, loss of non-interlayer porosity and evolution of pore structure, random-walk tortuosity, diffusion and stiffness. Predicted diffusion agrees with compacted Na-bentonite measurements.

\end{abstract}



\end{frontmatter}

\section{Introduction}
\label{sec1}

Bentonite is widely used in geotechnical and environmental engineering because of its low permeability and strong sealing capacity~\citep{pusch2015bentonite,murray2006bentonite}. These properties arise primarily from montmorillonite (MMT), a 2:1 layered phyllosilicate whose negative layer charge is balanced by exchangeable interlayer cations such as Na\textsuperscript{+}, Ca\textsuperscript{2+} and Mg\textsuperscript{2+}~\citep{marouf2021study,booker2004barrier,odom1984smectite}. Its high specific surface area, swelling capacity and cation-exchange capacity underpin applications in landfill liners, engineered barriers and adsorbents~\citep{murray2006bentonite,clem1961industrial}.

Compacted bentonite is also used as an engineered barrier in deep geological repositories (DGRs) for used nuclear fuel~\citep{murray2006bentonite,grambow2016geological,sellin2013use}. Sodium-rich MMT (Na--MMT) is often favoured because it swells more strongly and forms larger diffuse double layers than calcium variants~\citep{muhammad2022calcium,muhammad2019stabilization,sarkar2017preliminary,siddiqua2011evaluation}. Because groundwater velocities are extremely low under DGR conditions, mass transport is predominantly diffusion-controlled~\citep{allen1988bentonite,neuzil1986groundwater}. Diffusion coefficients depend strongly on dry density, microstructure and exchangeable cation~\citep{choi1996diffusive,muurinen2004ion,molera2002diffusion}. These variables redistribute water between interlayer and non-interlayer pores, which have markedly different transport properties~\citep{molera2003anion}, while the same evolving pore architecture controls hydro-mechanical behaviour~\citep{muurinen2013bentonite}.

Resolving this pore architecture is a multiscale challenge. X-ray diffraction (XRD), small-angle scattering (SAXS and SANS) and transmission electron microscopy (TEM) provide insight into nanostructure, but the intermediate mesostructure from nanometres to micrometres remains poorly resolved. In particular, these methods cannot capture the dynamic collective rearrangements of clay particles at aggregate scales~\citep{muurinen2013bentonite, cadars2012new,holmboe2012porosity,ferrage2018influence, devineau2006situ}. Atomistic simulations can resolve discrete hydration states and oscillatory interlayer forces through potentials of mean force (PMFs)~\citep{pedram2025investigating, zhang2022coarse, brochard2021swelling}, but remain restricted to scales below those relevant to macroscopic transport and stiffness. Traditional colloidal models such as Derjaguin--Landau--Verwey--Overbeek (DLVO) theory describe clay stability at separations of several nanometres and larger~\citep{derjaguin1941theory, verwey1947theory}, but do not capture the oscillatory hydration forces that arise during crystalline swelling at small interlayer spacings~\citep{pashley1984molecular}.

Platelet-based mesoscale models bridge this gap by preserving platelet anisotropy while representing aggregate-scale microstructural evolution and treating water and ions implicitly through effective interactions~\citep{ebrahimi2016mesoscale, ebrahimi2014mesoscale, ebrahimi2016effect, ghazanfari2022coarse, kang2020wettability, zhang2022coarse, zhang2023mechanical, zhang2025interlayer,le2023mesoscale, lin2024rheology, tong2025microscopic, tong2024micro, tong2024microscale, zhu2025mesoscale, shen2025coarse}. Such models access the length scales needed to resolve pore connectivity and platelet rearrangement and to extract transport and mechanical properties from simulated microstructures. PMFs obtained from all-atom molecular dynamics using the ClayFF force field~\citep{cygan2004molecular} describe the free-energy landscape as a function of platelet separation and therefore provide a physical basis for these effective interactions~\citep{pedram2025investigating}. Retaining the associated energy minima is essential because discrete swelling states determine interlayer spacing and influence the partitioning of water between pore environments.

Existing coarse-grained (CG) models balance computational efficiency against their ability to reproduce these hydration features. The Gay--Berne potential, for example, is an anisotropic extension of the Lennard--Jones potential that efficiently captures orientation-dependent interactions and platelet alignment~\citep{gay1981modification}. Its smooth interaction profile, however, cannot represent the multi-well energetics of crystalline swelling~\citep{ebrahimi2014mesoscale}. \citet{masoumi2017interparticle} fitted an analytical potential directly to atomistic PMFs, recovering oscillatory interactions, but the model is orientation-agnostic and computationally demanding~\citep{zhu2022potential}.

\citet{zhang2022coarse,zhang2023mechanical} subsequently represented hydrated Na--MMT platelets as assemblies of centre and edge interaction sites governed by Morse potentials augmented with three Gaussian terms per interaction type. The model reproduced structural and mechanical properties in agreement with experiments~\citep{zhang2023mechanical} and was later applied to transport by extracting porosity, pore-size distributions and tortuosity from three-dimensional reconstructions~\citep{zhang2023mechanical}. However, it does not fully recover the three-water (3--W) interlayer state observed in atomistic simulations and inferred from X-ray diffraction; the corresponding slit width is largely absent from its reconstructed pore-size distributions. The relative abundance of 1--W, 2--W and 3--W states determines interlayer spacing, the partitioning of interlayer and non-interlayer porosity and, consequently, the porosity--tortuosity combinations governing effective diffusion. Experiments likewise show that transport in Na-bentonite is highly sensitive to hydration state and microstructural connectivity~\citep{bourg2010connecting,bourg2015self,churakov2011up,appelo2013review,oscarson1992diffusion,sato2005effects,muurinen2013bentonite}. Moreover, the model relies on manually specified correction terms, which may limit its adaptability and transferability across clay types and configurations~\citep{zhang2022coarse,zhu2022potential}.

Here, we retain the site-based CG representation of \citet{zhang2022coarse}, which accommodates different platelet geometries and could be extended to flexible platelets, and introduce a hybrid interaction model combining a Morse baseline with a Gaussian process regression (GPR) correction. The Morse term captures the dominant short-range interaction, while the GPR term learns residual hydration-induced oscillations through a physics-informed kernel, avoiding a fixed set of manually prescribed correction functions. Implemented as a single tabulated potential, the model is trained on atomistic PMFs and evaluated across platelet geometries, layer-charge variants and configurations excluded from training. We then use it to connect interparticle energetics to pore structure, effective diffusion and quasi-static elastic response across compacted Na--MMT microstructures.

Section~\ref{sec2} describes the atomistic PMFs, CG representation, Morse+GPR framework and training procedure. Section~\ref{sec3} evaluates the learned interactions and applies the model to pore structure, accessible porosity, random-walk tortuosity, diffusion and elastic properties. We conclude with limitations and possible extensions to other ion--MMT systems and flexible platelets.

\section{Methods}
\label{sec2}

We construct the physics-informed CG model in three stages. First, atomistic molecular dynamics (MD) provides PMFs between hydrated Na--MMT platelets, which define the reference free-energy profiles for coarse-graining~\citep{pedram2025investigating, rodney2011modeling}. Second, these profiles are mapped onto the site-based platelet representation of~\citet{zhang2022coarse}, and a hybrid Morse--GPR pair potential is fitted for centre--centre, edge--centre and edge--edge interactions. Third, the fitted interactions are tabulated and used to generate compacted Na--MMT assemblies over a range of dry densities. The resulting structures are analysed for pore-size distribution, accessible porosity, random-walk tortuosity and elastic response.

\subsection{Reference atomistic interaction energies}

We use PMFs from our previous atomistic study as reference data for coarse-graining; full simulation details are reported by \citet{pedram2025investigating}. A PMF describes the free-energy variation along a selected reaction coordinate~\citep{trzesniak2007comparison, kastner2011umbrella, sprik1998free}. Here, the coordinate is the separation between two clay structures, and the PMF is obtained by integrating the mean force required to maintain successive separations using constrained dynamics~\citep{sprik1998free}. Between separations $r_a$ and $r_b$:
\begin{equation}
\Delta E_{a-b}=\int_{r_a}^{r_b}F_r\,dr,
\label{eq:one}
\end{equation}
where $F_r$ is the mean constraint force at separation $r$, with positive and negative values denoting repulsion and attraction, respectively.

Simulations were performed with LAMMPS~\citep{plimpton1995fast, thompson2022lammps} using ClayFF~\citep{cygan2004molecular}, which represents van der Waals interactions with Lennard--Jones potentials and electrostatics with Coulombic terms. Three Na--MMT structures were generated using ATOM\\ \mbox{~\citep{holmboe2019atom}:}
\begin{enumerate}
    \item a finite hexagonal platelet, $\mathrm{[Na_{8}][Si_{168}][Mg_{8}Al_{34}]O_{192}(OH)_{78}}$;
    \item a fully periodic sheet, $\mathrm{[Na_{26}][Si_{320}][Mg_{26}Al_{134}]O_{800}(OH)_{160}}$; and
    \item a semi-periodic sheet, periodic along $y$ and non-periodic along $x$,\\
    $\mathrm{[Na_{28}][Si_{336}][Mg_{28}Al_{140}]O_{824}(OH)_{200}}$.
\end{enumerate}
The hexagonal geometry minimizes broken silicate rings~\citep{lammers2017molecular, underwood2020large, zhang2022coarse}. Harmonic bond and angle terms represented edge O--H bonds and metal--O--H angles~\citep{pouvreau2019structure, thompson2022lammps, pouvreau2017structure}, and water was described by the flexible SPC model~\citep{berendsen1981interaction}.

Face-to-face and edge-to-edge arrangements were considered. Each configuration was solvated with pre-equilibrated water, and molecules within 2~\AA{} of any clay atom were removed to eliminate overlaps. Systems were equilibrated for 50~ps in the $NPT$ ensemble at 300~K and 1~atm, followed by 1~ns constrained $NVT$ simulations at 300~K with a 1~fs timestep. Constraint forces were sampled on selected octahedral metal atoms: 11 of 42 in the finite platelet, 12 of 160 in the fully periodic sheet and 14 of 168 in the semi-periodic sheet. Sampling multiple constrained atoms distributes the applied constraint across each structure while retaining a well-defined platelet separation.

The three structures sample finite, semi-periodic and fully periodic geometries as well as several layer-charge states arising from different numbers of isomorphic substitutions. To broaden the training set further, we include the fully periodic Na--MMT model of \citet{honorio2017hydration} \newline
$\mathrm{[Na_6^+][Si_{62}Al_2][Mg_4Al_{28}]O_{160}(OH)_{32}\cdot} n\mathrm{H_2O}$. Our Arizona Na--MMT structures represent layer charge through octahedral Mg$^{2+}$-for-Al$^{3+}$ substitution, whereas the Honorio model contains both octahedral Mg$^{2+}$-for-Al$^{3+}$ and tetrahedral Al$^{3+}$-for-Si$^{4+}$ substitutions. It therefore introduces a distinct layer chemistry and charge distribution for testing whether the fitted interactions generalize beyond a single substitution pattern. PMFs reported by \citet{zhang2022coarse} were reserved for out-of-set evaluation and were generated using the same constrained-dynamics procedure. Across all training and test configurations, Na--MMT surface charge densities range from approximately {\(-0.18\)} to
\(-0.26~\mathrm{C\,m^{-2}}\), consistent with reported smectite values~\citep{christidis2003determination,mahesh2011synthesis,bailey2015smectite}. 
Although surface charge density describes the overall magnitude of the layer charge, the local electrostatic environment also depends on substitution location. The combination of charge states, substitution patterns and platelet geometries therefore provides a more stringent assessment of transferability. Table~\ref{tab:pmf_cases} summarizes the training and test configurations.

\begin{table}[htbp]
\centering
\scriptsize
\setlength{\tabcolsep}{3pt}
\renewcommand{\arraystretch}{1.15}
\caption{PMF configurations used to train and test the Morse+GPR coarse-grained potential. The Na--MMT models differ in geometry, layer charge and substitution pattern.}
\label{tab:pmf_cases}
\vspace{2pt}
\begin{tabularx}{\textwidth}{@{}L{1.2cm} L{2.1cm} L{3.6cm} L{3.0cm} L{1.5cm}@{}}
\toprule
Set & Source & Mineral models & Arrangement & Distance range \\
\midrule
Training & \citet{pedram2025investigating}
& small hexagonal platelet--small hexagonal platelet
& face-to-face & $\sim$10--22~\AA{} \\
Training & \citet{pedram2025investigating}
& small hexagonal platelet--small hexagonal platelet
& edge-to-edge & $\sim$24--32~\AA{} \\
Training & \citet{honorio2017hydration}
& fully periodic sheet--fully periodic sheet
& face-to-face & $\sim$10--22~\AA{} \\
Training & \citet{pedram2025investigating}
& fully periodic sheet--small hexagonal platelet
& face-to-face & $\sim$10--22~\AA{} \\
\midrule
Testing & \citet{pedram2025investigating}
& semi-periodic sheet--semi-periodic sheet
& face-to-face & $\sim$10--24~\AA{} \\
Testing & \citet{zhang2022coarse}
& fully periodic sheet--small hexagonal platelet
& face-to-edge & $\sim$16--26~\AA{} \\
Testing & \citet{zhang2022coarse}
& small hexagonal platelet--large hexagonal platelet
& face-to-face & $\sim$10--18~\AA{} \\
Testing & \citet{zhang2022coarse}
& small hexagonal platelet--small hexagonal platelet
& face-to-face, 30$^\circ$ in-plane rotation
& $\sim$10--36~\AA{} \\
\bottomrule
\end{tabularx}
\end{table}

\subsection{Coarse-grained platelet representation}

We adopt and extend the site-based representation of \citet{zhang2022coarse} to preserve platelet geometry and hydration-sensitive interaction environments at reduced computational cost. Each platelet contains two rigid layers of CG sites separated by 5.44~\AA{}, corresponding to the thickness of one T--O--T layer in the minimized atomistic structure. The upper layer is offset in-plane by 2.835~\AA{}, producing the observed $\sim61^\circ$ tilt. This two-layer construction preserves the sheet architecture and provides a scaffold that could later be extended to platelet bending, shear or delamination. Centre sites form the platelet backbone and represent interactions away from hydroxylated edges, whereas edge sites describe Si--OH and Al/Mg--OH environments and their associated hydrogen-bonding contributions. Centre and edge spacings are approximately 6.5 and 5.5~\AA{}, respectively, reflecting the silicate-ring diameter and edge-hydroxyl spacing. Sites are arranged in a hexagonal pattern to minimize broken silicate rings and to provide a uniform mapping for finite, semi-periodic and fully periodic structures~\citep{lammers2017molecular, underwood2020large}.

\begin{figure}[t!]
    \centering
    \includegraphics[width=0.85\textwidth]{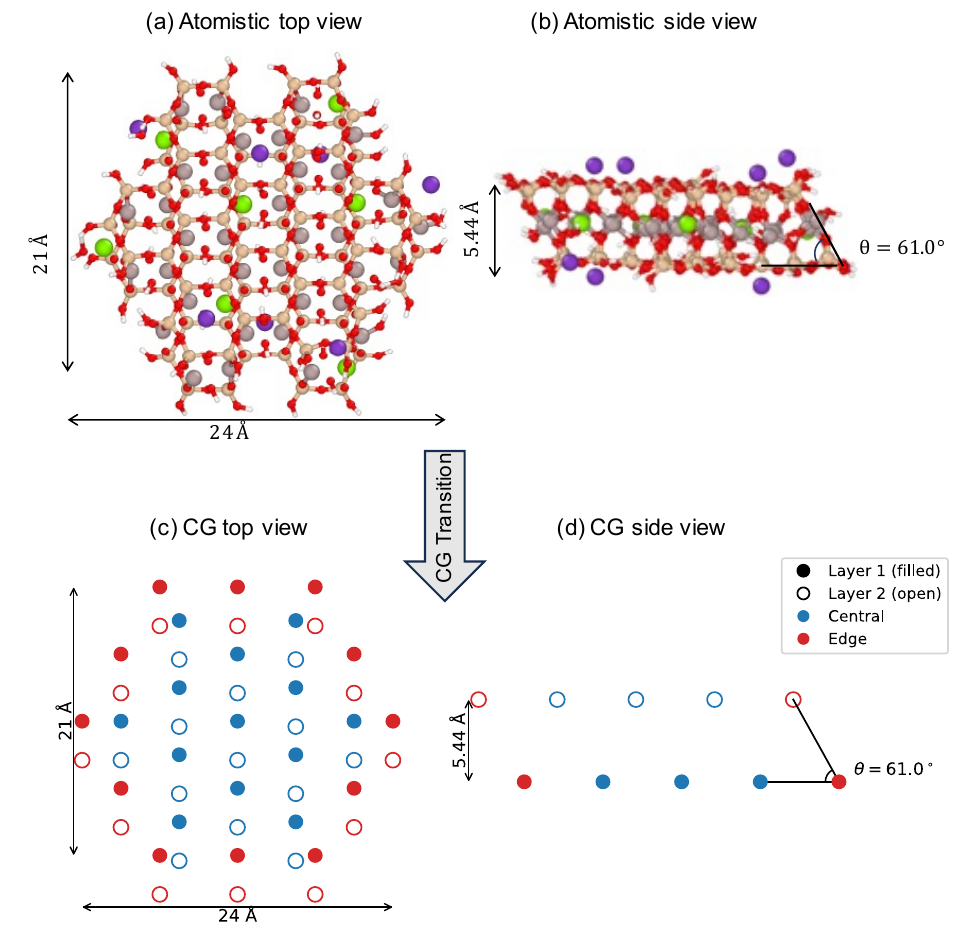}
    \caption{Atomistic-to-coarse-grained mapping of a small hexagonal Na--MMT platelet. (a,b) Atomistic top and side views. (c) CG top view with centre (blue) and edge (red) sites; filled and open markers denote the lower and upper layers. (d) CG side view showing the 5.44~\AA{} layer separation and $\sim61^\circ$ tilt generated by the lateral offset.}
    \label{fig:1_ex}
\end{figure}

For example, the 482-atom finite platelet is reduced to 50 CG sites while retaining its overall dimensions and distinct centre and edge environments (Fig.~\ref{fig:1_ex}). Water and ions are not represented explicitly; their free-energy contribution is embedded in the effective interactions fitted to the atomistic PMFs. Pairwise distances between sites on different platelets are the only structural descriptors entering the potential. Distances are evaluated to a radial cutoff of 25~\AA{}, which is required to include the longest edge--edge contributions. If a periodic cell is smaller than the cutoff, it is replicated before neighbour-list construction so that all relevant periodic images are included without duplicate interactions.

\subsection{Hybrid pair-potential formulation}

\begin{figure}[t]
    \centering
    \includegraphics[width=0.4\textwidth]{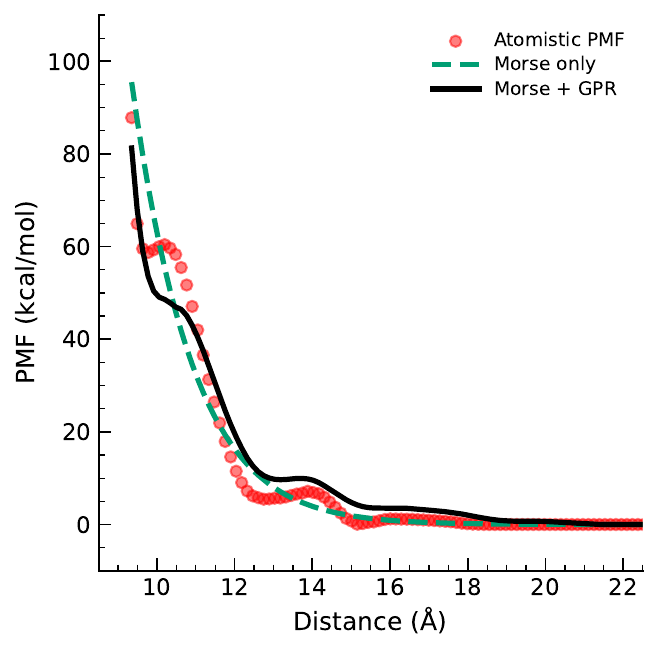}
    \caption{Morse+GPR decomposition of a fully periodic Na--MMT sheet--sheet PMF. The Morse term captures the baseline trend, and the GPR correction recovers the hydration-induced oscillations and multiple minima.}
    \label{fig:morse_gpr_decomposition}
\end{figure}

For centre--centre (CC), edge--centre (EC) and edge--edge (EE) pairs, the interaction is
\begin{equation}
U_{\text{pair}}^{(\tau)}(r)
=U_{\text{Morse}}^{(\tau)}(r)+U_{\text{GPR}}^{(\tau)}(r),
\qquad \tau\in\{\mathrm{CC},\mathrm{EC},\mathrm{EE}\}.
\label{eq:pair}
\end{equation}
The Morse baseline~\citep{morse1929diatomic}
\begin{equation}
U_{\text{Morse}}^{(\tau)}(r)
=D_e^{(\tau)}\left[
 e^{-2\alpha^{(\tau)}(r-r_e^{(\tau)})}
-2e^{-\alpha^{(\tau)}(r-r_e^{(\tau)})}
\right]
\label{eq:morse}
\end{equation}
contains the well depth $D_e^{(\tau)}$, equilibrium distance $r_e^{(\tau)}$ and stiffness parameter $\alpha^{(\tau)}$. The Morse term supplies an interpretable repulsive core and baseline attractive trend for each interaction type. It does not, however, reproduce the secondary minima and longer-range oscillations produced by discrete hydration states. These residual features are represented by the GPR correction, as illustrated for a fully periodic sheet--sheet PMF in Fig.~\ref{fig:morse_gpr_decomposition}.

The atomistic PMF is a total configuration energy, whereas the CG model is expressed as pair interactions. For configuration $m$ at platelet separation $d_m$, the Morse contribution is therefore summed over all interacting site pairs:
\begin{equation}
E_{\mathrm{Morse}}(d_m)=
\sum_{\tau\in\{\mathrm{CC},\mathrm{EC},\mathrm{EE}\}}
\sum_{(i,j)\in\tau}
U_{\mathrm{Morse}}^{(\tau)}\!\big(r_{ij}(d_m)\big),
\label{eq:morse_total}
\end{equation}
where $r_{ij}(d_m)$ is the corresponding site separation. The configuration-level residual is
\begin{equation}
\delta E(d_m)=E_{\mathrm{PMF}}(d_m)-E_{\mathrm{Morse}}(d_m),
\label{eq:deltaE_def}
\end{equation}
and is decomposed into type-specific pair corrections:
\begin{equation}
\delta E(d_m)\approx
\sum_{\tau\in\{\mathrm{CC},\mathrm{EC},\mathrm{EE}\}}
\sum_{(i,j)\in\tau}
\delta U^{(\tau)}\!\big(r_{ij}(d_m)\big).
\label{eq:deltaE_decomp}
\end{equation}
The GPR contribution in Eq.~\eqref{eq:pair} is identified with this pairwise residual, $U_{\mathrm{GPR}}^{(\tau)}(r)\equiv\delta U^{(\tau)}(r)$. The correction is not fitted as a separate function for each atomistic PMF.
Instead, the same CC, EC and EE pairwise GPR residual functions are used for
all configurations.
For each configuration, the Morse and GPR pair contributions are both summed
over the corresponding site pairs to obtain the predicted total PMF energy at
each separation.


Each correction is expanded over $S$ Gaussian kernels with shared support locations $\{s_k\}_{k=1}^{S}\subset[0,r_c]$:
\begin{equation}
U_{\text{GPR}}^{(\tau)}(r)=
\sum_{k=1}^{S}\alpha_k^{(\tau)}
\tilde{\kappa}\!\left(r;s_k,\ell^{(\tau)}\right),
\label{eq:gpr}
\end{equation}
where
\begin{equation}
\tilde{\kappa}(r;s_k,\ell^{(\tau)})=
\begin{cases}
\exp\!\left[-\dfrac{(r-s_k)^2}{2(\ell^{(\tau)})^2}\right]-\exp\!\left[-\dfrac{9}{2}\right],
& |r-s_k|\leq3\ell^{(\tau)},\\[8pt]
0,& |r-s_k|>3\ell^{(\tau)}.
\end{cases}
\label{eq:kernel}
\end{equation}
Subtracting $\exp(-9/2)$ makes each Gaussian basis function vanish continuously at its local cutoff, $|r-s_k|=3\ell^{(\tau)}$. The separate global pair cutoff is fixed at $r_c=25~\AA{}$ when constructing pair-distance lists. The support locations $\{s_k\}$ are shared among CC, EC and EE so that all interaction types use the same radial basis, whereas the coefficients $\alpha_k^{(\tau)}$ and Gaussian widths $\ell^{(\tau)}$ remain interaction-specific. A single width is used for all supports within each interaction type, controlling overall smoothness without introducing a separate width for every basis function. The cutoff is fixed, while the number and locations of support points, coefficients and widths are determined during fitting. This sparse kernel representation can reproduce multiple oscillatory wells from a limited set of PMFs while retaining smooth pair potentials~\citep{deringer2021gaussian, chen2021gaussian}.

\subsection{Parameter fitting and regularization}

The potential was fitted in PyTorch~\citep{paszke2019pytorch} using Adam~\citep{adam2014method} for the GPR contributions and limited-memory Broyden--Fletcher--Goldfarb--Shanno optimization~\citep{liu1989limited} for the Morse parameters. The three interaction types were first fitted sequentially to reduce parameter coupling. CC parameters were determined from fully periodic sheet--sheet PMFs, where only centre interactions contribute~\citep{honorio2017hydration}. EC parameters were then fitted to PMFs for a fully periodic sheet interacting with a finite platelet while holding CC fixed. Finally, EE parameters were fitted to edge-to-edge and stacked finite-platelet configurations with both CC and EC fixed.

For each interaction type, fitting proceeded from the simplest baseline to the full flexible model:
\begin{enumerate}
    \item \textbf{Morse baseline.} The GPR term was disabled, and $(D_e,\alpha,r_e)$ were fitted to the overall PMF shape and short-range repulsion.
    \item \textbf{Residual correction.} GPR coefficients and the interaction-specific width were fitted using support points uniformly distributed on $[0,25]~\AA{}$, with their locations fixed.
    \item \textbf{Joint fixed-support refinement.} Morse parameters \((D_e, \alpha, r_e)\), GPR coefficients \(\alpha_k\), and
length scale \(\ell\) were jointly optimized while the support-point
locations \(s_k\) were kept fixed.

    \item \textbf{Global adaptive refinement.} All interaction types and model parameters were optimized jointly. Support locations and the total support number were also adjusted, with points added, removed or relocated in high-residual regions to improve the balance among configurations and limit overfitting.
\end{enumerate}
Stages 1--3 were applied in the order CC $\rightarrow$ EC $\rightarrow$ EE, freezing previously fitted terms; stage 4 provided global refinement.

The baseline objective was
\begin{equation}
\mathcal{L}_{\mathrm{MSE}}=
\sum_i w_{\mathrm{rel}}(E_i)(\hat{E}_i-E_i)^2
+\lambda_1\!\sum_{k=S-9}^{S}|\alpha_k|,
\label{eq:loss_mse}
\end{equation}
where $E_i$ and $\hat{E}_i$ are the reference and predicted PMF energies at spacing $d_i$. The energy-dependent relative weight
\begin{equation}
w_{\mathrm{rel}}(E)=
\min\left[\max\left(\frac{1}{|E|+\epsilon},w_{\min}\right),w_{\max}\right]
\label{eq:energy_weight}
\end{equation}
emphasizes low-energy regions, including hydration minima, while the clipping bounds prevent a small number of near-zero energies from dominating the fit. We used $\epsilon=10^{-3}$, $w_{\min}=1/90$ and $w_{\max}=1$. An \(L_1\) regularization term was applied only to the ten largest-radius GPR
coefficients, with \(\lambda_1 = 0.01\).
This promotes sparsity in the tail and encourages the residual correction to decay at long range. All GPR coefficients were initialized to zero so that early optimization began from the stable Morse baseline.

After the global weighted-MSE fit, selected spacing ranges with strongly
oscillatory residuals were further refined with a Huber objective. In this
local-refinement step, the index \(i\) in Eq.~(12) runs only over the PMF data
points within the selected spacing ranges, rather than over the full PMF range
used in Eq.~(9).
Defining
\[
\Delta\tilde{E}_i=\frac{\hat{E}_i-E_i}{E_{\mathrm{scale}}},
\qquad E_{\mathrm{scale}}=\max_j|E_j|,
\]
we use
\begin{equation}
\mathcal{L}_{\mathrm{Huber}}(\Delta\tilde{E}_i)=
\begin{cases}
\dfrac{1}{2}(\Delta\tilde{E}_i)^2,&|\Delta\tilde{E}_i|\leq\delta,\\[4pt]
\delta\left(|\Delta\tilde{E}_i|-\dfrac{1}{2}\delta\right),&|\Delta\tilde{E}_i|>\delta,
\end{cases}
\label{eq:huber}
\end{equation}
with $\delta=10^{-2}$. The corresponding refinement objective was
\begin{equation}
\mathcal{L}_{\mathrm{GPR}}=
\sum_i w_{\mathrm{rel}}(E_i)
\mathcal{L}_{\mathrm{Huber}}(\Delta\tilde{E}_i)
+\lambda_1\!\sum_{k=S-9}^{S}|\alpha_k|.
\label{eq:loss_huber}
\end{equation}
The weighted-MSE objective was used first to establish a smooth global fit. Where the amplitude or phase of strongly oscillatory hydration features remained under-resolved, the GPR parameters were re-optimized over the affected spacing ranges using the Huber objective. Normalizing by the largest absolute energy in the corresponding PMF prevents datasets with larger energy scales from dominating this refinement. The Huber form also reduces sensitivity to individual highly weighted points while retaining the emphasis on low-energy regions.

During global refinement, support-point ordering and the $[0,25]~\AA{}$ bounds were enforced through a log-space parameterization of inter-point spacings. Short-range supports were fixed to stabilize the repulsive region, while the remaining locations were updated by gradient-based optimization. A discrete adaptation step inserted or relocated supports where residuals remained large, combining differentiable parameter fitting with adaptive basis selection. The initial learning rate was 0.5 and was reduced when the loss plateaued; subsequent interaction-specific refinements used stage-dependent optimizer settings. Final Morse parameters, support locations, widths, GPR coefficients and tabulated CC, EC and EE potentials are provided in the Supplementary Material. These files fully define the interaction model used below and permit direct implementation without repeating the fitting procedure.

\subsection{Assembly preparation and compaction}

For mesoscale simulations, the fitted Morse+GPR energies were evaluated on a 1000-point radial grid from 0.5 to 25~\AA{} and converted to tabulated energies and forces compatible with the LAMMPS table-potential format. A Savitzky--Golay filter was applied to the total energy profiles before force tabulation to suppress small numerical irregularities and ensure stable interpolation. The resulting tables are available in the \href{https://github.com/yaldapedram/nammt-morse-gpr-cg}{online repository}.

Two ensembles, each containing 1000 platelets, were generated to examine the effects of platelet-size distribution and loading geometry. The first comprised anisotropic polydisperse systems with diameters sampled from the experimental 10--50~nm distribution~\citep{zhang2025interlayer}. The second comprised nominally isotropic systems in which every platelet had a diameter of 12~nm. Platelets were placed randomly in periodic cells using overlap checks to prevent steric clashes. In the polydisperse systems, normals were assigned random in-plane rotations and out-of-plane tilts up to $\pm20^\circ$ to facilitate subsequent compaction along a preferred direction. Orientations in the 12-nm systems were fully random to preserve isotropy.

Five independent 12-nm systems were initialized in cubic boxes with side lengths of 800, 1000, 1200, 1400 and 1600~\AA{}. These different starting volumes reduce dependence on a single packing rather than defining different final densities. Systems were annealed for 50~ns in the $NVT$ ensemble at a fictitious thermostat temperature of 1000~K, cooled to 300~K over 50~ns, and equilibrated at 300~K and 1~atm in the $NPT$ ensemble. This produced reference structures near $0.8~\mathrm{g\,cm^{-3}}$ with average box lengths of approximately 650~\AA{}. The thermostat temperature is a numerical annealing parameter because the CG interactions were parameterized from PMFs at 300~K.

The same five replicas were carried through sequential $NPT$ compression at 5, 10, 30, 100 and 200~atm, producing dry densities from 0.9 to $1.3~\mathrm{g\,cm^{-3}}$ and providing replicate statistics at each density. Polydisperse systems followed the same thermal schedule but were compressed in the $NP_zT$ ensemble, with pressure applied along $z$ to represent uniaxial compaction. Representative configurations at 0.8 and $1.3~\mathrm{g\,cm^{-3}}$ are provided in the Supplementary Material. The final structures were equilibrated within the selected CG preparation and compaction protocols, but they should not be interpreted as unique equilibrium fabrics of real smectite. Clay microstructure is history-dependent, and alternative annealing or loading paths may yield different pore structures and derived properties.

\subsection{Pore geometry and accessible porosity}

Void geometry was evaluated by Monte Carlo insertion using PorosityPlus as a post-processing method~\citep{opletal2018porosityplus}. Each equilibrated configuration was treated as a static assembly of rigid, impenetrable CG spheres with effective radius 3.5~\AA{}, consistent with the atomistic-to-CG mapping. For each configuration, $4\times10^6$ probe insertions were attempted. At every accepted point in the void phase, 300 biased random walks, each limited to 100 steps with a maximum step length of 2.0~\AA{}, searched for the largest sphere that could be inscribed locally without intersecting a platelet. Accessible porosity $\theta$ was calculated as the ratio of accepted to attempted insertions. The coordinates and maximal diameters of accepted points were retained to construct pore-size distributions and three-dimensional pore visualizations.

The normalized pore-size distribution $PSD(d)$ is defined from the local maximal probe diameter $d$ such that
\[
\int PSD(d)\,\mathrm{d}d=\theta.
\]
Thus, the area under each distribution equals the probe-accessible porosity.

\subsection{Random-walk tortuosity}

The accessible pore geometry was voxelized on a $500\times500\times500$ grid and analysed using constrained lattice random walks implemented with Pytrax~\citep{tranter2019pytrax}. Each voxel was assigned a value between 0 and 1 that determines the probability of entering it: 0 represents solid clay, 1 represents fully accessible bulk-like pore space and intermediate values represent interlayer regions with hindered mobility.

Three idealized interlayer treatments were considered to separate geometric confinement from local mobility effects: (i) an interlayer-forbidden case in which interlayer voxels were impermeable, (ii) an interlayer-throttled case in which they remained accessible but with reduced move probabilities, and (iii) a no-throttle case in which interlayers were treated as fully open pore space. In the throttled case, 1-, 2- and 3-W regions were assigned diffusion factors of 0.05, 0.27 and 0.44 relative to bulk water, respectively, based on long-time Na-smectite self-diffusion ratios~\citep{bourg2010connecting}. Interlayer shells were classified by distance from the nearest platelet surface, following \citet{zhang2025interlayer} and evidence that local diffusivity varies with surface distance~\citep{rotenberg2010molecular}. Voxels beyond approximately three water diameters were assigned unit accessibility.

For each configuration, 10{,}000 walkers were initialized uniformly among voxels with non-zero accessibility. Each trajectory comprised 5000 attempted moves to one of the six nearest neighbours under periodic boundary conditions. A move into a zero-valued voxel was rejected and the walker remained in place; otherwise, the move was accepted with a probability equal to the target voxel value. Selected moves advanced the walker by one voxel, while rejected moves still contributed to elapsed simulation time. Random-walk tortuosity $\tau_{\mathrm{RW}}$ was obtained from the inverse slope of the mean-square-displacement--time relation. In the no-throttle and interlayer-forbidden cases, $\tau_{\mathrm{RW}}$ reflects restrictions imposed by the connected pore geometry. In the throttled case, it additionally includes the prescribed reduction in interlayer mobility and is therefore not a purely geometric path-length ratio. Walker number and trajectory length were increased until further increases did not change the ensemble mean within statistical uncertainty.

\subsection{Quasi-static elastic response}
\label{methodMechanicalproperties}

Elastic constants were determined from linear stress--strain responses under small quasi-static deformations~\citep{zhang2022coarse,zhang2023mechanical,ebrahimi2014mesoscale}. Each equilibrated configuration was cooled gently to 0.01~K and held in the $NVT$ ensemble to suppress thermal stress fluctuations while retaining the inter-platelet interactions parameterized at 300~K. Six independent loading modes were applied: uniaxial strain along $x$, $y$ and $z$, and simple shear in the $xy$, $xz$ and $yz$ planes. For each mode, a total strain of \(0.08\%\) was accumulated over
approximately 53~ns in increments of approximately
{\(0.0015\%\)}. After every increment, the structure was relaxed for 1.2~ns and the Cauchy stress was sampled for a further 0.4~ns. Stress--strain curves were linear over this interval, and their least-squares slopes defined the corresponding stiffness elements.

For the anisotropic layered polydisperse systems, single-valued Young's and shear moduli were calculated from the full stiffness tensor using Voigt--Reuss--Hill averaging, enabling comparison with experimental modulus--void-ratio data~\citep{voigt1889ueber, reuss1929berechnung, hill1952elastic}. The stiffness tensor was first symmetrized and inverted to obtain the compliance tensor. Voigt and Reuss bounds were then calculated for the bulk and shear moduli and combined using the Hill average. For the nominally isotropic 12-nm systems, the tensor was close to the conventional isotropic form, from which Young's and shear moduli were obtained directly.

\section{Results}
\label{sec3}

We first assess the Morse+GPR interactions against atomistic PMFs and then use the resulting CG assemblies to quantify pore structure, tortuosity, diffusion scaling and mechanical response.

\subsection{Morse+GPR interaction model}

The benchmark combines PMFs from the present atomistic calculations and prior studies and spans platelet geometry, boundary conditions, orientation and layer charge~\citep{zhang2022coarse,honorio2017hydration,pedram2025investigating}. Face-to-face and edge-to-edge configurations of fully periodic sheets and small hexagonal platelets were used for training. Held-out tests comprise a semi-periodic face-to-face configuration, a face-to-edge configuration, a size-mismatched platelet pair and a face-to-face pair rotated by \(30^\circ\).

\begin{figure}[!t]

  \centering
  \includegraphics[width=0.65\textwidth]{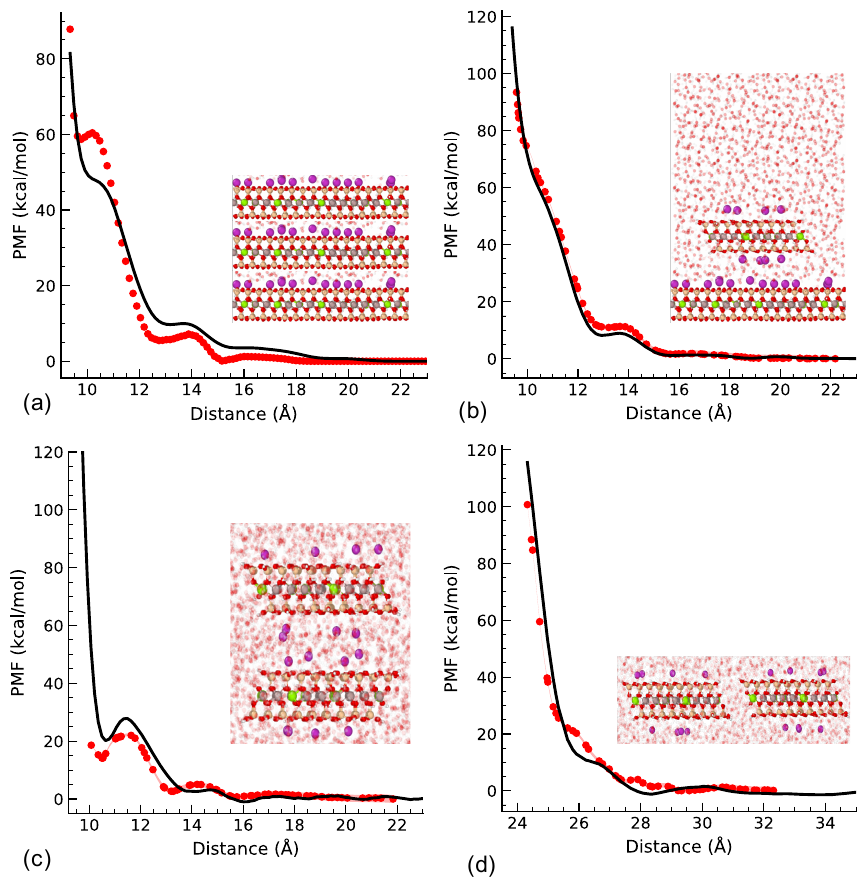}
  \caption{Training-set PMFs used to fit the Morse+GPR coarse-grained interactions. Red symbols denote the atomistic reference PMFs; the red shaded band indicates the uncertainty associated with the atomistic PMFs. Black curves show the corresponding Morse+GPR predictions. Insets illustrate the training geometries: (a) fully periodic sheet vs.\ fully periodic sheet, face-to-face; (b) fully periodic sheet vs.\ small hexagonal platelet, face-to-face; (c) small hexagonal vs.\ small hexagonal platelet, face-to-face; and (d) small hexagonal vs.\ small hexagonal platelet, edge-to-edge.}

  \label{fig:TrainingPMFs}
\end{figure}

\begin{figure}[!t]

  \centering
  \includegraphics[width=0.65\textwidth]{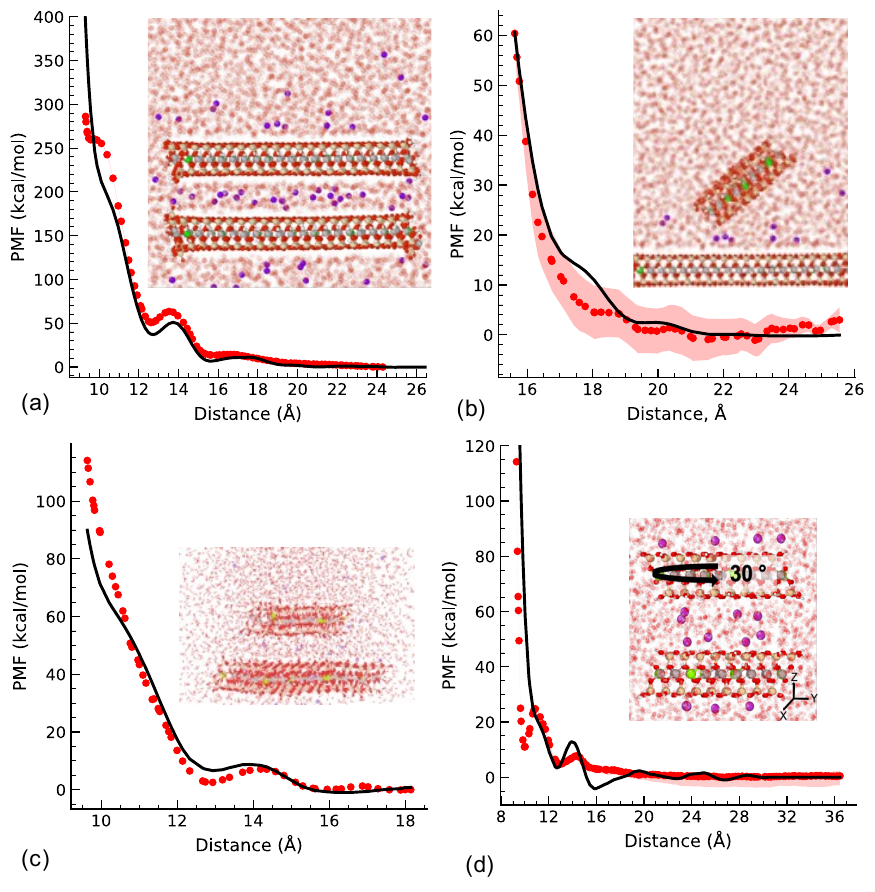}
  \caption{Predicted versus reference PMFs for held-out Na--MMT configurations. Red symbols show the reference (atomistic) PMFs and the red shaded band indicates the corresponding uncertainties. Black curves show the coarse-grained Morse+GPR predictions. Insets illustrate the corresponding geometries. (a) Semi-periodic sheet vs.\ semi-periodic sheet, face-to-face. (b) Fully periodic sheet vs.\ small hexagonal platelet, face-to-edge. (c) Small hexagonal vs.\ large hexagonal platelet, face-to-face. (d) Small hexagonal vs.\ small hexagonal platelet, face-to-face with a $30^\circ$ in-plane rotation.}

  \label{fig:TestingPMFs}
\end{figure}

The fitting targets were the short-range repulsion and the positions and depths of the hydration minima. The minima determine the free-energy differences among hydration states and the equilibrium interlayer spacings. These spacings, in turn, control how pore volume is partitioned between interlayers and larger voids and how those regions connect through the microstructure. Because interlayer and non-interlayer pores have distinct diffusivities, errors in the hydration minima would propagate directly into porosity, tortuosity and transport predictions~\citep{bourg2006tracer,appelo2013review}. Figures~\ref{fig:TrainingPMFs} and~\ref{fig:TestingPMFs} compare the CG and atomistic PMFs.

Across the training set (Fig.~\ref{fig:TrainingPMFs}), the model reproduces near-contact repulsion and the observed 0-W, 1-W, 2-W and 3-W minima. The agreement is important in both energy and separation: the barrier heights determine the accessibility of neighbouring states, while the positions of the wells determine the preferred slit widths. The model also captures the sequence of hydration transitions reported in atomistic and experimental studies~\citep{antognozzi2001observation,clark1937study,israelachvili1983molecular,pashley1984molecular}.

The largest training residuals occur for systems with different layer charges and structural provenance. 
The small platelets have a surface charge density of {\(-0.20~\mathrm{C\,m^{-2}}\)}, compared with
{\(-0.26~\mathrm{C\,m^{-2}}\)} for the periodic system.
Layer charge alters interlayer ion--water structure, hydration-well depths and the local PMF slopes that determine swelling pressure~\citep{seppala2016effect}. Differences in isomorphic substitution and sampling protocol also contribute: the periodic PMFs include Wyoming Na--MMT and GCMC data, whereas the platelet systems use Arizona Na--MMT and constrained MD. Published PMFs likewise vary with clay provenance, geometry, platelet size and sampling method~\citep{zhu2022potential}. The residuals should therefore be interpreted relative to this underlying variability rather than as deviations from a unique universal PMF.

The held-out tests (Fig.~\ref{fig:TestingPMFs}) retain the principal physical features:

\textbf{1) Semi-periodic sheet--sheet, face-to-face.}
The CG curve closely follows the atomistic PMF across the full range, including the 0-W-to-1-W transition and long-range decay. The minima occur at nearly the correct positions and depths, with only small systematic residuals. This is the strongest held-out result because it shows transfer from the finite and fully periodic training geometries to a distinct semi-periodic boundary condition.

\textbf{2) Periodic sheet--small platelet, face-to-edge.}
The overall profile and spacing dependence are reproduced, but energies are overestimated at \(16\text{--}20~\text{\AA}\) and the first minimum is too shallow. The test system has fewer interlayer Na\(^{+}\) ions and a lower layer charge than the corresponding training configurations. These changes alter the first hydration well and the short-range barrier through their effect on ion--water organization and swelling pressure~\citep{seppala2016effect}. Because charge is not an explicit descriptor, the potential represents an average over the training conditions. It consequently captures the broad PMF trend but smooths the small high-frequency oscillations beyond the first minimum.

\textbf{3) Small--large platelets, face-to-face.}
The predicted minima occur at nearly the correct separations and the phase of the oscillations is retained. The model slightly underestimates repulsion at \(d\approx9.5\text{--}11~\text{\AA}\) and produces a shallower second well near \(12\text{--}13~\text{\AA}\), while the two curves converge beyond \(16~\text{\AA}\). A size-mismatched pair has a greater relative contribution from corners and edges than the configurations used to isolate the face interactions. Local charge patches, nearby Na\(^{+}\) ions and structured edge water can therefore strengthen the atomistic short-range response. Because the EC and EE terms depend only on separation, they average over edge termination, corner geometry and hydration environment, softening the repulsive rise without substantially shifting the minima.

\textbf{4) Small platelet pair, \(30^\circ\) face-to-face rotation.}
The model preserves the oscillation spacing and long-range decay but overestimates the short-range barrier and weakens the first well. The second and third minima remain close to the atomistic positions, and their depths show no consistent bias. Rotation changes the registry of opposing surface sites and the packing of the first hydration layers, reducing the atomistic barrier during close approach. The pairwise CG interactions contain no descriptor of this site matching and instead encode an orientation-averaged response. The resulting error is therefore concentrated at the shortest separations, while the intermediate- and long-range behaviour remains transferable.

Overall, one tabulated two-body potential reproduces the main hydration barriers and wells and transfers across boundary conditions, size mismatch, orientation and layer-charge variants without case-specific refitting~\citep{zhang2022coarse}. The predicted hydration spacings and barriers are consistent with ranges reported for layered silicates~\citep{israelachvili1983molecular,pashley1984molecular}, while charge-dependent shifts and held-out residuals are comparable to published effects of layer charge and inter-study variability~\citep{seppala2016effect,zhu2022potential}. The model also avoids prescribing a fixed number or location of analytical correction terms and can be refined over selected separation ranges without changing the overall framework. Its principal limitation is the absence of explicit orientation and edge-environment descriptors. Distinct corner, kink and straight-edge sites are represented by the same distance-dependent terms, so their effects are recovered only in an averaged sense. Edge-type descriptors or a limited three-body correction could address these residuals while retaining the tabulated implementation.

\subsection{Transport properties}

\subsubsection{Pore-size distribution}

\begin{figure}[htbp]
\captionsetup[subfigure]{font=normalsize}
  \centering

  \begin{subfigure}{0.47\textwidth}
    \centering
    \begin{overpic}[width=\linewidth]{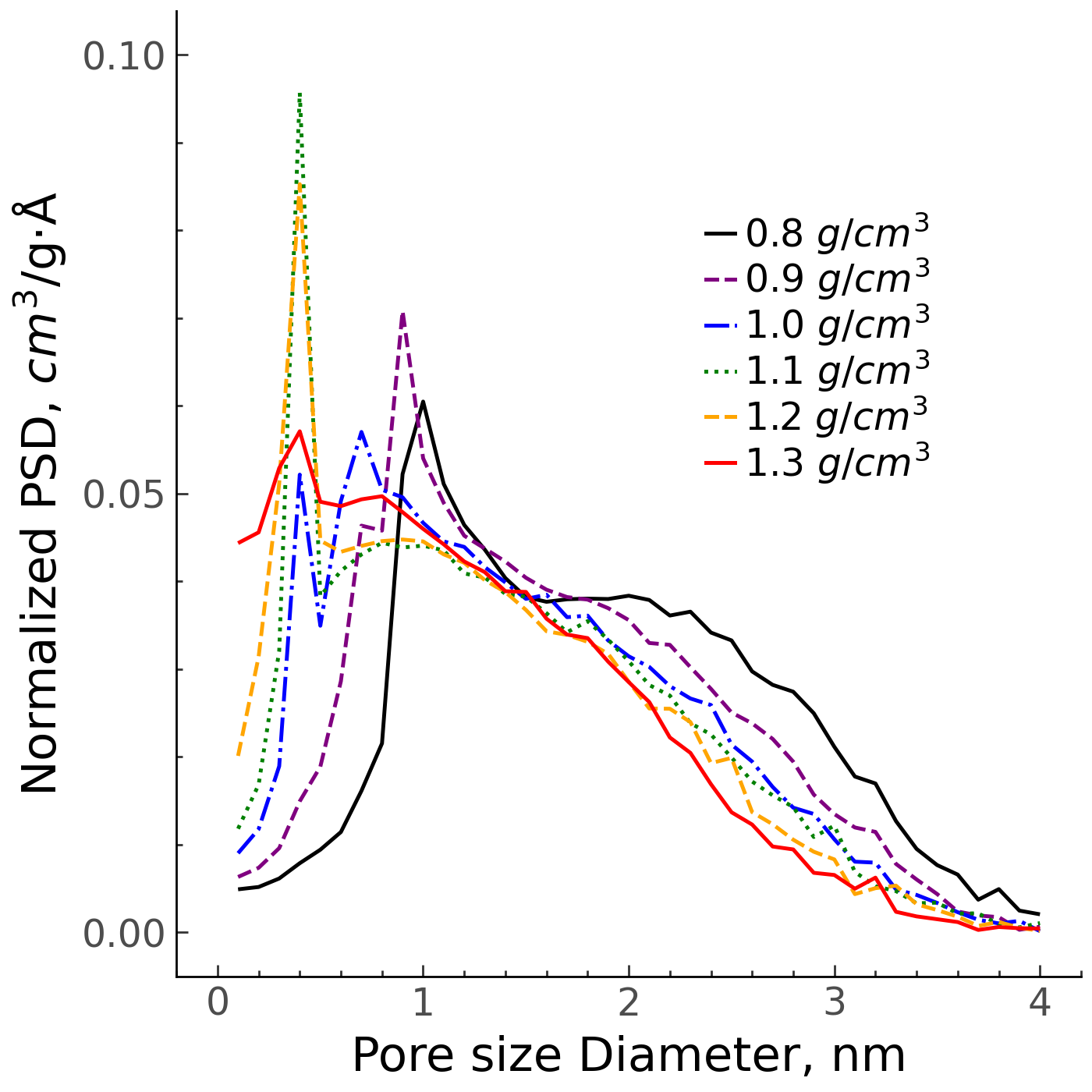}
      \put(20,99){\textbf{12~nm isotropic systems}}
      \put(3,5){\normalsize (a)}
    \end{overpic}
    \label{fig:psd_iso}
  \end{subfigure}
  \hfill
  \begin{subfigure}{0.47\textwidth}
    \centering
    \begin{overpic}[width=\linewidth]{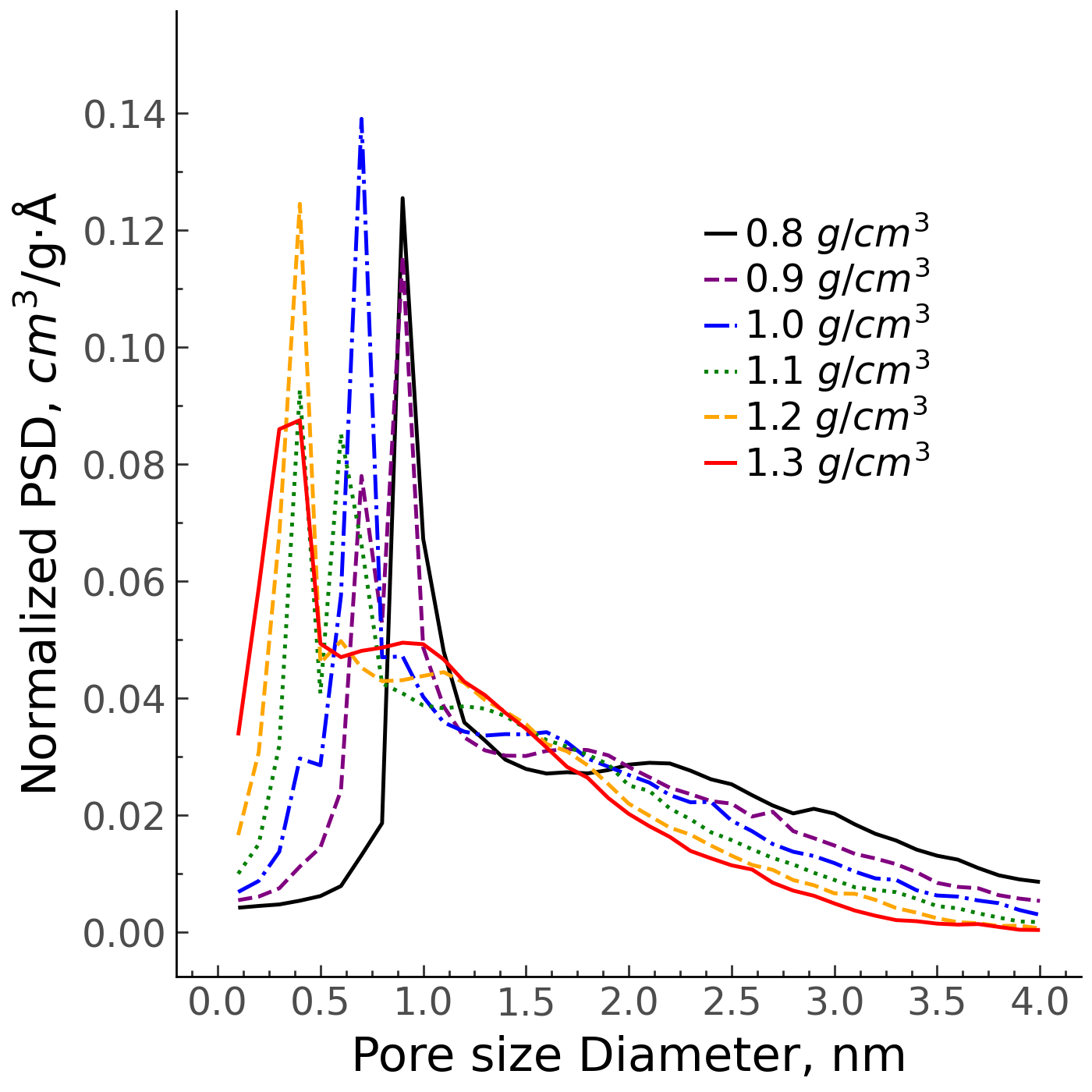}
      \put(19,99){\textbf{Polydisperse NP\(_z\)T systems}}
      \put(3,5){\normalsize (b)}
    \end{overpic}
    \label{fig:psd_poly}
  \end{subfigure}

  \caption{Normalized pore–size distributions of tracer–accessible void space in Na--MMT at different dry densities for (a) 12~nm isotropic and (b) polydisperse NP\(_z\)T systems.}
  \label{fig:psd}
\end{figure}

\begin{figure}[!t]
    \centering
    \begin{overpic}[width=0.95\textwidth]{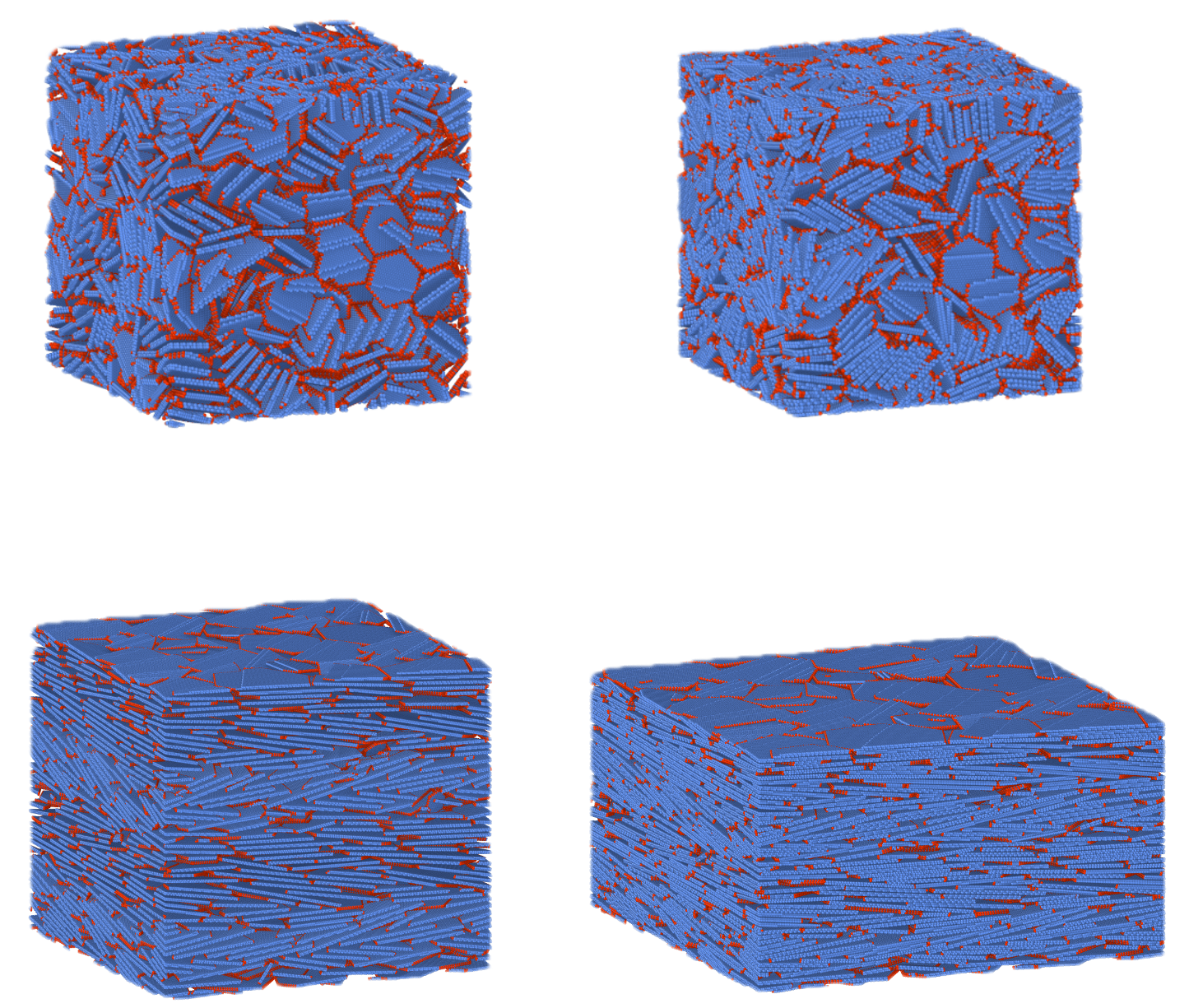}

        \put(49, 90){\makebox(0,0){\textbf{Isotropic systems}}}

        \put(18, 45){\makebox(0,0){\small (a) $\rho_d=0.8~\mathrm{g\,cm}^{-3}$}}
        \put(72, 45){\makebox(0,0){\small (b) $\rho_d=1.3~\mathrm{g\,cm}^{-3}$}}

        \put(49, 39){\makebox(0,0){\textbf{Polydisperse NP\(_z\)T systems}}}

        \put(18, -2){\makebox(0,0){\small (c) $\rho_d=0.8~\mathrm{g\,cm}^{-3}$}}
        \put(72, -2){\makebox(0,0){\small (d) $\rho_d=1.3~\mathrm{g\,cm}^{-3}$}}

    \end{overpic}
    \vspace{16pt} 
    \caption{CG microstructures used for the PSD analysis: (a,b) isotropic 12-nm systems and (c,d) polydisperse $z$-compressed systems, each at $\rho_d=0.8$ and $1.3~\mathrm{g\,cm^{-3}}$.}
    \label{fig:microstructure-snapshots}
\end{figure}

Figure~\ref{fig:psd} shows the PSDs from \(0.8\) to \(1.3~\mathrm{g\,cm^{-3}}\) for the isotropic and polydisperse systems. Both contain sharp interlayer modes near \(0.40\), \(0.75\) and \(1.00~\mathrm{nm}\), associated with 1-W, 2-W and 3-W states, and a broad tail from larger pores between stacks and aggregates.

Compaction shifts the dominant interlayer population from 3-W toward 1-W. At \(0.8~\mathrm{g\,cm^{-3}}\), the \(1.0~\mathrm{nm}\) mode dominates. A weaker \(0.75~\mathrm{nm}\) shoulder appears by \(0.9~\mathrm{g\,cm^{-3}}\), and the two wider hydration states have comparable weight near \(1.0~\mathrm{g\,cm^{-3}}\). From \(1.1\) to \(1.3~\mathrm{g\,cm^{-3}}\), the PSD collapses progressively toward the \(0.40~\mathrm{nm}\) mode. Simultaneously, the non-interlayer tail contracts, indicating closure of voids between stacks and aggregates and redistribution of accessible volume toward interlayers. The same sequence appears in both microstructures, but the polydisperse systems show sharper interlayer peaks and a smaller large-pore tail. Their anisotropic packing therefore produces a narrower distribution of local hydration environments, consistent with the more layered microstructures in Fig.~\ref{fig:microstructure-snapshots}.

This combination of discrete interlayer peaks, a broad larger-pore population and its density dependence agrees qualitatively with SAXS, NMR and chloride-exclusion analyses of compacted Na-bentonite and earlier CG reconstructions~\citep{muurinen2013bentonite,zhang2023mechanical,zhang2025interlayer}. In particular, the resolved 3-W mode is consistent with experimental interpretations of face-to-face Na--MMT stacks~\citep{muurinen2013bentonite,holmboe2012porosity}. The 2-W-to-3-W transition represents approximately a 20\% increase in basal spacing and a 50\% increase in interlayer pore width, so recovering this state materially changes the available pore volume rather than merely adding a minor spectral feature. Its presence addresses the under-representation of the 3-W minimum in earlier CG interactions and allows the density-driven redistribution among all three hydration states to emerge in the mesoscale assemblies. The peak positions also agree with slit widths inferred for discrete hydration states from X-ray diffraction profile modelling~\citep{holmboe2011free,holmboe2012porosity}.

\subsubsection{Accessible pore volume}

The absolute total porosity provides an upper bound on the connected pore volume. Assuming a rigid, incompressible solid,
\[
\theta_{\mathrm{total}}^{\mathrm{abs}}
=
\frac{V_{\mathrm{cell}}-V_{\mathrm{solid}}}{V_{\mathrm{cell}}}
=
1-\frac{\rho_d}{\rho_s},
\]
where \(\rho_s=2.75~\mathrm{g\,cm^{-3}}\) is the grain density used by \citet{muurinen2013bentonite}. This upper bound includes disconnected, occluded and interlayer-confined volume. Table~\ref{tab:porosity-summary} additionally reports \(\theta_{\mathrm{acc}}\), the connected interlayer contribution \(\theta_{\mathrm{IL}}\), and \(f_{\mathrm{IL}}=\theta_{\mathrm{IL}}/\theta_{\mathrm{acc}}\).

For context, the table includes measured water porosity and modelled non-interlayer porosity from Muurinen \emph{et al.}~\citep{muurinen2013bentonite}. Their pore partition was inferred from NMR, SAXS and microstructural modelling rather than direct geometric segmentation. In particular, their interlayer quantity includes water in isolated stacks and dead-end galleries, whereas \(\theta_{\mathrm{IL}}\) here includes only connected, probe-accessible pathways. The values are therefore complementary rather than definition-matched: the former measures total interlayer water content, while the latter represents the interlayer fraction of a percolating transport network. Isolated galleries may exchange locally with neighbouring pores but contribute little to long-range diffusion.

In both microstructures, \(\theta_{\mathrm{acc}}\) decreases from approximately \(0.63\) at \(0.8~\mathrm{g\,cm^{-3}}\) to \(0.39\) at \(1.3~\mathrm{g\,cm^{-3}}\). The polydisperse systems are lower by \(0.01\text{--}0.03\) at a given density, consistent with more efficient packing under \(z\)-directed compression. Meanwhile, \(f_{\mathrm{IL}}\) increases from \(4.1\%\) to \(34.8\%\) in the isotropic systems and from \(1.3\%\) to \(34.3\%\) in the polydisperse systems. The difference between fabrics is therefore greatest at low density, where packing controls whether interlayers belong to the connected network, but nearly disappears after substantial compaction. Between \(1.2\) and \(1.3~\mathrm{g\,cm^{-3}}\), the connected interlayer volume is nearly constant while total connected porosity continues to fall. The rising interlayer fraction at the highest densities is thus driven mainly by loss of non-interlayer connectivity rather than continued creation of interlayer volume.

Relative to \(\theta_{\mathrm{total}}^{\mathrm{abs}}\), connected porosity is \(11\text{--}14\%\) lower at \(0.8~\mathrm{g\,cm^{-3}}\) and \(25\text{--}28\%\) lower at \(1.3~\mathrm{g\,cm^{-3}}\). The widening gap indicates that compaction creates a growing fraction of occluded, disconnected or strongly confined volume. At \(1.0\) and \(1.3~\mathrm{g\,cm^{-3}}\), the simulated values (\(0.521\text{--}0.527\) and \(0.379\text{--}0.394\)) are also below the measured water porosities (\(0.64\) and \(0.53\)). This is expected in part because the experiments include water-filled regions that need not belong to a percolating geometric network. Finite voxel resolution may further smooth narrow throats and rough microvoids that remain accessible to small experimental probes~\citep{muurinen2013bentonite}. The small difference between the two simulated fabrics is therefore less important than the systematic distinction between total water-filled and connected probe-accessible porosity. Overall, compaction shifts the connected network from larger inter-aggregate pores toward persistent interlayer pathways.

\begin{table}[!t]
\centering
\caption{Porosity summary of isotropic 12-nm and polydisperse $z$-compressed CG systems. For each dry density, the absolute total porosity $\theta_{\mathrm{total}}^{\mathrm{abs}}$ is calculated as $1-\rho_d/\rho_s$, using the grain density $\rho_s=2.75~\mathrm{g\,cm^{-3}}$ reported by \citet{muurinen2013bentonite}, and is used as an upper bound. For each CG microstructure, $\theta_{\mathrm{acc}}$ is the connected accessible porosity, $\theta_{\mathrm{IL}}$ is the connected interlayer contribution and $f_{\mathrm{IL}}=\theta_{\mathrm{IL}}/\theta_{\mathrm{acc}}$ (\%). The final two columns give the measured water porosity and modelled non-interlayer porosity from \citet{muurinen2013bentonite}.}
\label{tab:porosity-summary}
\footnotesize
\setlength{\tabcolsep}{5pt}
\renewcommand{\arraystretch}{1.15}
\begin{tabular*}{\textwidth}{@{\extracolsep{\fill}} c c ccc ccc cc @{}}
\toprule
\shortstack{Dry density \\ $(\mathrm{g\,cm^{-3}})$} &
$\theta_{\mathrm{total}}^{\mathrm{abs}}$ &
\multicolumn{3}{c}{Isotropic 12-nm} &
\multicolumn{3}{c}{Polydisperse $z$-compressed} &
\multicolumn{2}{c}{Muurinen \textit{et al.}} \\
\cmidrule(lr){3-5} \cmidrule(lr){6-8} \cmidrule(l){9-10}
& & $\theta_{\mathrm{acc}}$ & $\theta_{\mathrm{IL}}$ & $f_{\mathrm{IL}}$ (\%) 
  & $\theta_{\mathrm{acc}}$ & $\theta_{\mathrm{IL}}$ & $f_{\mathrm{IL}}$ (\%) 
  & $\theta_{\mathrm{water}}$ & $\theta_{\mathrm{non\text{-}IL}}$ \\
\midrule
0.7 & 0.745 & --    & --    & --   & --    & --    & --   & 0.75 & 0.48 \\
0.8 & 0.709 & 0.628 & 0.026 &  4.1 & 0.613 & 0.008 &  1.3 & --   & --   \\
0.9 & 0.673 & 0.579 & 0.050 &  8.6 & 0.563 & 0.031 &  5.5 & --   & --   \\
1.0 & 0.636 & 0.527 & 0.097 & 18.4 & 0.521 & 0.065 & 12.5 & 0.64 & 0.34 \\
1.1 & 0.600 & 0.490 & 0.127 & 25.9 & 0.471 & 0.105 & 22.3 & --   & --   \\
1.2 & 0.564 & 0.450 & 0.141 & 31.3 & 0.422 & 0.130 & 30.7 & --   & --   \\
1.3 & 0.527 & 0.394 & 0.137 & 34.8 & 0.379 & 0.130 & 34.3 & 0.53 & 0.21 \\
1.6 & 0.418 & --    & --    & --   & --    & --    & --   & 0.42 & 0.11 \\
\bottomrule
\end{tabular*}
\end{table}

\subsubsection{Tortuosity}

The isotropic assemblies yield a scalar tortuosity \(\tau_{\mathrm{iso}}\), whereas the polydisperse systems yield \(\tau_\perp\) and \(\tau_\parallel\), normal and parallel to the mean platelet orientation. Table~\ref{tab:tortuosity} reports the full-network, IL-throttled and IL-forbidden cases. These are random-walk tortuosities rather than geometric path-length ratios. Porosity is treated separately in the diffusion scaling; in the throttled case, voxel-dependent move probabilities also enter \(\tau_{\mathrm{RW}}\), so the reported value includes both network geometry and the assigned interlayer mobility penalty.

For the isotropic systems, \(\tau_{\mathrm{iso}}\) rises monotonically from \(1.95\) to \(5.65\) over the simulated density range, reflecting the progressive removal of broad, direct pathways. It lies between the two directional values of the polydisperse full network and is closer to \(\tau_\perp\), consistent with an orientational average rather than either limiting direction. The order parameter remains \(S\simeq0.15\text{--}0.18\), confirming weak texture under the second-order Legendre definition~\citep{allen2017computer} and placing these systems near the low-orientation end of the range reported by Ferrage \emph{et al.}~\citep{ferrage2018influence}. The increase in tortuosity is therefore primarily a consequence of pore closure rather than strong platelet alignment.

The polydisperse full network is strongly anisotropic: \(\tau_\perp\) changes only from \(1.15\) to \(1.42\), whereas \(\tau_\parallel\) increases from about \(9\) to \(24\). Accordingly, \(\tau_\parallel/\tau_\perp\) rises from approximately \(8\) to \(17\). Within this microstructure, transport parallel to the mean platelet plane must traverse increasingly long quasi-planar corridors, while the available cross-layer connections remain comparatively direct. The larger order parameter, \(S\simeq0.32\text{--}0.33\), approximately twice the isotropic value, indicates a moderately oriented fabric rather than either an isotropic or nearly perfectly aligned assembly~\citep{ferrage2018influence}.

Interlayer treatment has little influence at low density, consistent with the small interlayer fraction of the connected pore volume, but becomes important as compaction closes the larger channels. At \(1.3~\mathrm{g\,cm^{-3}}\), throttling raises \(\tau_\perp\) from \(1.42\) to \(2.82\) and \(\tau_\parallel\) from \(23.8\) to \(40.8\). The IL-forbidden values remain above the full-network values but below the throttled values at high density. This ordering is not paradoxical: removing the most resistive interlayer corridors reroutes walkers through fewer but higher-conductance off-plane connections, whereas retaining those corridors with very low mobility permits repeated entry into quasi-trapping regions. The difference between the two scenarios therefore measures both the availability of interlayer paths and the penalty associated with traversing them.

\begin{table}[!t]

\centering
\caption{Random-walk tortuosity versus dry density. 
\(\tau_{\mathrm{iso}}\) is the scalar random-walk tortuosity of the isotropic 12-nm aggregates. 
For the polydisperse \(z\)-compressed systems, \(\tau_\perp\) and \(\tau_\parallel\) denote
random-walk tortuosities normal and parallel to the mean platelet orientation, respectively, for three
connectivity scenarios: (i) full pore network accessible, 
(ii) interlayer voxels assigned a reduced local diffusive conductance, and 
(iii) interlayer voxels treated as non-permeable.}
\label{tab:tortuosity}
\footnotesize
\setlength{\tabcolsep}{3pt}
\renewcommand{\arraystretch}{1.08}

\begin{tabular*}{\textwidth}{@{\extracolsep{\fill}}cccccccc@{}}
\toprule
\shortstack{Dry density \((\mathrm{g\,cm^{-3}})\)} &
\(\tau_{\mathrm{iso}}\) &
\multicolumn{2}{c}{Full network} &
\multicolumn{2}{c}{IL-throttled} &
\multicolumn{2}{c}{IL-forbidden} \\
\cmidrule(lr){3-4}\cmidrule(lr){5-6}\cmidrule(l){7-8}
&
&
\(\tau_\perp\) & \(\tau_\parallel\) &
\(\tau_\perp\) & \(\tau_\parallel\) &
\(\tau_\perp\) & \(\tau_\parallel\) \\
\midrule
0.8 & 1.953 & 1.148 &  9.169 & 1.178 &  9.913 & 1.179 & 10.277 \\
0.9 & 2.290 & 1.167 & 11.307 & 1.273 & 13.559 & 1.272 & 13.733 \\
1.0 & 2.968 & 1.188 & 13.589 & 1.461 & 17.675 & 1.403 & 17.502 \\
1.1 & 3.529 & 1.256 & 16.721 & 1.800 & 24.262 & 1.607 & 22.559 \\
1.2 & 4.310 & 1.331 & 19.990 & 2.304 & 32.226 & 1.877 & 28.857 \\
1.3 & 5.649 & 1.421 & 23.812 & 2.823 & 40.767 & 2.185 & 35.130 \\
\bottomrule
\end{tabular*}
\end{table}

\subsubsection{Diffusion scaling}

Diffusion through compacted bentonite is commonly represented using accessible porosity and tortuosity~\citep{briggs2017multi,oscarson1992diffusion,oscarson1994surface,choi1996diffusive}. The same framework has been applied to weakly sorbing anions~\citep{oscarson1992diffusion,sato2005effects} and tritiated water~\citep{sato1992effect,sato1992study,kato1994estimation,nakazawa1999activation,sato2003fundamental,garcia2004diffusion,sato2005effects,sanchez2008self,bourg2015self,suzuki2004study,kozaki1999effect,torikai1996study}. We therefore use the standard scaling~\citep{oscarson1992diffusion,oscarson1994surface,appelo2013review}:

\begin{equation}
  \frac{D}{D_0}
  =
  \frac{\theta^{(s)}}{\tau_{\mathrm{RW}}^{(s)}},
  \label{eq:diff-scaling}
\end{equation}

where \(D\) and \(D_0\) are the effective and bulk-water diffusion coefficients, respectively, and \(s\) denotes the full-network, IL-throttled or IL-forbidden scenario. The porosity and tortuosity must correspond to the same scenario because blocking interlayers changes both the connected volume and the available trajectories. Here, \(\tau_{\mathrm{RW}}^{(s)}\) is obtained from the slope of the random-walk mean-square displacement and already represents the ratio of free-space to porous-medium diffusivity. It therefore enters directly rather than as a squared geometric path-length factor~\citep{tranter2019pytrax}.

\begin{figure}[htbp]
    \centering
    \includegraphics[width=0.8\textwidth]{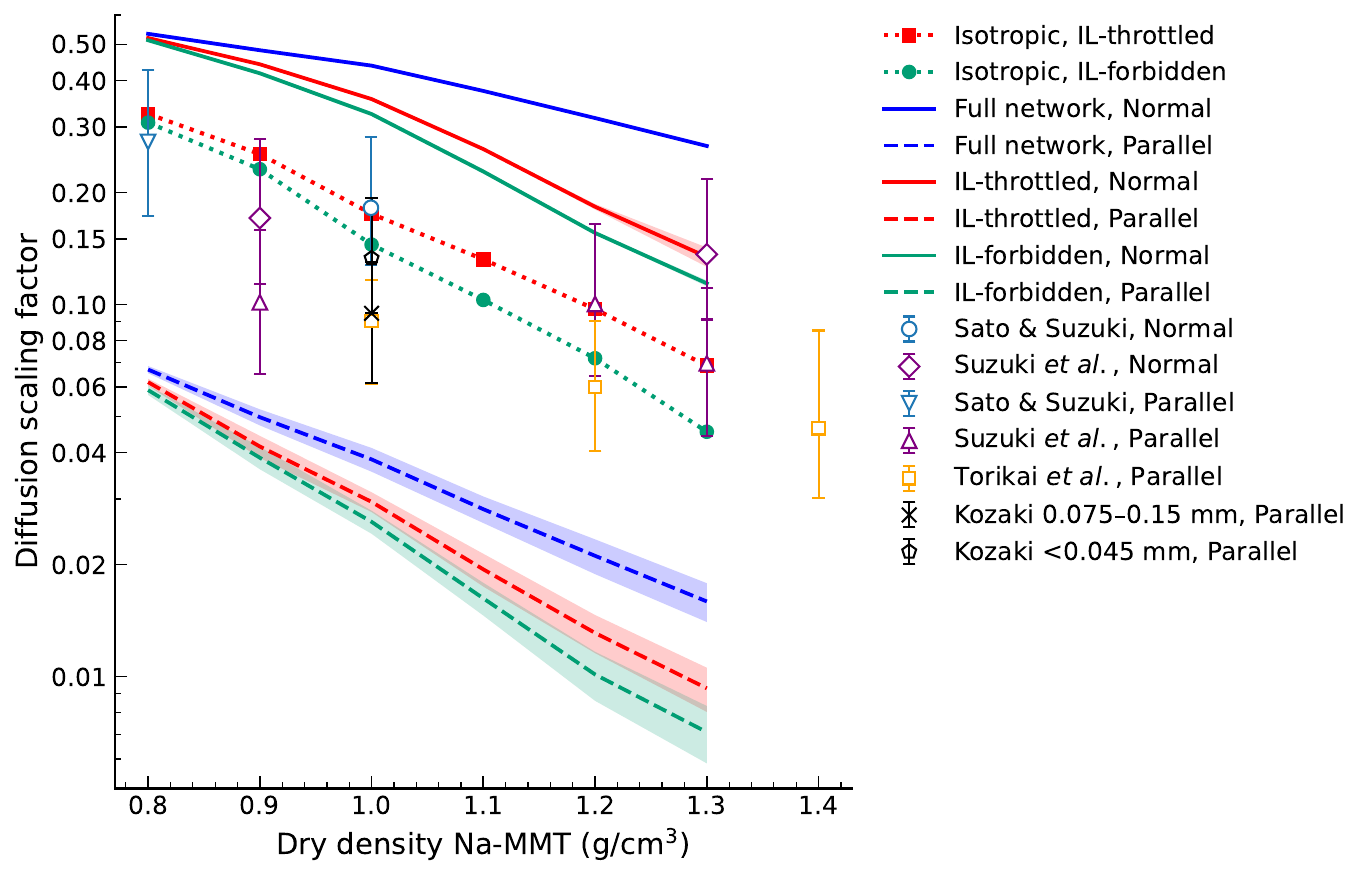} 
    \caption{Diffusion scaling factor \(D/D_0=\theta^{(s)}/\tau_{\mathrm{RW}}^{(s)}\) for a neutral tracer in Na-MMT as a function of dry density. 
Red and green dotted curves show predictions for the 12~nm isotropic systems for the IL-throttled and IL-forbidden cases; these predictions are direction-independent. 
Solid and dashed lines show the corresponding normal- and parallel-to-compaction components for the polydisperse NP\(_z\)T systems under three interlayer treatments (full network, IL-throttled, IL-forbidden); shaded bands denote $\pm$ one standard error across the five simulations at each density.
Symbols with error bars denote experimental tritiated-water diffusion coefficients measured normal and parallel to the compaction direction in compacted Kunipia-F Na-bentonite (\(\sim 98\%\) Na--MMT) ~\citep{sato2003fundamental,suzuki2004study,torikai1996study,kozaki1999effect}.}
    \label{fig:diffusion}
\end{figure}

Figure~\ref{fig:diffusion} compares the predicted neutral-tracer scaling with tritiated-water measurements for compacted Kunipia-F. Because the isotropic systems yield one direction-averaged value while the experiments report directions normal and parallel to compaction, the two experimental series provide a natural envelope. The isotropic IL-throttled and IL-forbidden predictions lie within this envelope and reproduce the strong decline with density. This decline follows directly from decreasing \(\theta^{(s)}\), increasing \(\tau_{\mathrm{RW}}^{(s)}\), closure of larger pores and the shift from 3-W toward 1-W interlayers. The pore cross-section available to a tracer therefore decreases at the same time as the random-walk resistance increases.

For the polydisperse systems, predicted diffusion normal to compression is generally above the corresponding data, while the parallel component is below it. The simulated fabric is therefore more anisotropic than the experimental samples. Uniaxial preparation and rigid platelets favour ordered stacks and relatively persistent directional pathways. Disorder, bending, warping and aggregate-scale defects in real bentonite would interrupt these paths and reduce the contrast between directions. The anisotropic offset is consequently more plausibly associated with the idealized preparation and platelet mechanics than with the hydration minima of the interaction potential. Despite this offset, both microstructures reproduce the magnitude and density dependence of the experimental scaling without fitting to diffusion data.

\subsection{Mechanical response}
\label{sec:mech_results}
\begin{figure}[htbp]
  \centering
  \includegraphics[width=0.95\linewidth]{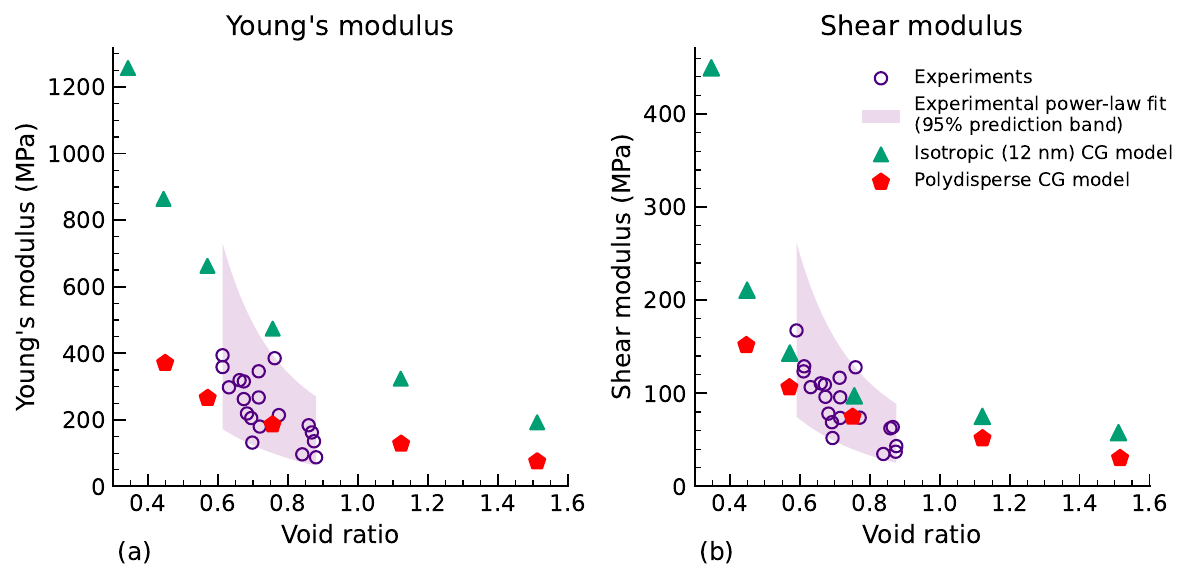}
  \caption{
  Young's modulus and shear modulus of compacted Na--MMT as a function of void ratio.
  Open circles show experimental Na-bentonite data ~\citep{pusch1983stress, borgesson2006earthquake, kiviranta2011quality, eloranta2012experimental, hancilova2016coupled, dixon2019review}; the shaded region denotes the 95\% prediction
  band of a power-law fit to the experimental points. Triangles indicate the isotropic
  (12~nm) CG model, and pentagons indicate the polydisperse CG model. 
  The left panel shows Young's modulus, and the right panel shows shear modulus.
  }
  \label{fig:voidratio_moduli}
\end{figure}

Figure~\ref{fig:voidratio_moduli} compares \(E\) and \(G\) with experimental modulus--void-ratio data. For the isotropic 12-nm systems, \(E\) is systematically high at a given void ratio, whereas most \(G\) values lie within the experimental prediction band. Their monodisperse, comparatively regular fabric lacks the broad platelet-size distribution, packing defects and microcracks present in real bentonite. These idealizations are expected to affect axial stiffness more strongly than the averaged shear response, consistent with the different levels of agreement for \(E\) and \(G\).

The polydisperse systems are softer: both moduli lie substantially closer to the experimental envelope, with several intermediate-void-ratio points within the reported scatter. This improvement is obtained without changing the interaction potential and therefore isolates the influence of platelet-size distribution and packing. Larger platelets, small platelets and the voids generated by their imperfect packing provide additional deformation modes that are absent from a monodisperse assembly. Much of the over-stiffness of the 12-nm systems consequently appears to arise from microstructural idealization rather than the pair potential itself. The Morse+GPR interactions combined with an experimentally informed size distribution produce effective moduli broadly compatible with laboratory measurements.

The stiffness tensors also reflect the imposed fabrics. In the isotropic series, \(C_{11}\), \(C_{22}\) and \(C_{33}\), and likewise \(C_{44}\), \(C_{55}\) and \(C_{66}\), remain similar at low and intermediate densities. A mild directional contrast develops with compaction, consistent with some preferred stacking, but the response remains close to isotropic. The polydisperse systems show clearer transverse isotropy: \(C_{11}\) and \(C_{22}\) remain comparable, while \(C_{33}\) and the shear components involving deformation across the layers differ more strongly. This pattern is expected for a layered fabric produced by one-dimensional compression. Given the finite CG boxes, limited replicas and small deformation intervals, individual tensor components should be interpreted as qualitative trends rather than calibrated predictions. Nevertheless, the emergence of nearly isotropic and transversely isotropic responses in the corresponding microstructures supports the sensitivity of the model to fabric.

\section{Discussion and conclusions}
\label{sec4}

We developed a physics-informed CG interaction model for hydrated Na--MMT that combines a Morse baseline with a GPR correction trained on atomistic PMFs. The Morse term provides an interpretable representation of short-range repulsion and attraction, while the GPR correction captures the oscillatory, multi-well free-energy landscape associated with discrete hydration states. The resulting interactions are implemented as a single tabulated potential, allowing atomistically informed platelet interactions to be used efficiently in mesoscale simulations.

Across the training configurations, the model captures the locations and general relative depths of the 0--W, 1--W, 2--W and 3--W minima, including the 3--W state that has been under-represented in previous closed-form CG potentials~\citep{zhang2023mechanical, zhang2022coarse, zhang2025interlayer}. The held-out tests further show transfer across the geometries, boundary conditions, orientations and layer-charge variants examined here. Residual errors are concentrated at short separations in asymmetric, rotated and size-mismatched configurations. These discrepancies are consistent with the use of distance-only EC and EE interactions, which average over differences in local charge, edge termination, corner geometry and orientation. Nevertheless, the model generally preserves the ordering and spacing of hydration minima, the barriers between adjacent states and the long-range decay of the PMFs. The tests therefore support the use of a shared tabulated potential for mesoscale simulations while also identifying where additional local or orientational descriptors would be most valuable.

The PMF-informed interactions generate a clear compaction-driven evolution of Na--MMT pore structure. In both the isotropic monodisperse and uniaxially compacted polydisperse assemblies, the PSDs contain distinct interlayer modes near 0.40, 0.75 and 1.0~nm, corresponding to 1--W, 2--W and 3--W states, together with a broader population of non-interlayer pores. Increasing dry density shifts the dominant interlayer population from 3--W toward 2--W and ultimately 1--W, while progressively eliminating larger pores. Accessible porosity consequently decreases, whereas the interlayer contribution rises to approximately one-third of the connected pore volume at the highest density. These trends are consistent with experimental microstructural characterizations and previous CG analyses of compacted Na--bentonite~\citep{sato1992effect,sato1992study,kato1994estimation,nakazawa1999activation,sato2003fundamental,garcia2004diffusion,sato2005effects,sanchez2008self,muurinen2013bentonite,zhang2023mechanical,zhang2025interlayer}. The resolution of a pronounced 3--W peak at low density, followed by its disappearance during compaction, further indicates that the learned PMFs place the hydration minima at separations consistent with experimentally inferred 1--W, 2--W and 3--W spacings~\citep{holmboe2011free,holmboe2012porosity}.

These structural changes translate directly into the predicted transport response. Random-walk tortuosity increases with dry density as connected non-interlayer pores close and transport becomes increasingly confined to narrow, indirect pathways. Combining tortuosity with tracer-accessible porosity through
\[
D/D_0=\theta^{(s)}/\tau_{\mathrm{RW}}^{(s)}
\]
reproduces the observed decrease in neutral-tracer diffusion with compaction. The isotropic 12~nm systems yield diffusion coefficients within the experimental range of tritiated-water measurements normal and parallel to the compaction direction. The polydisperse NP$_z$T systems reproduce the density dependence but overpredict diffusion normal to the preferred platelet orientation and underpredict it in the parallel direction~\citep{sato1992effect,sato1992study,kato1994estimation,nakazawa1999activation,sato2003fundamental,garcia2004diffusion,sato2005effects,sanchez2008self,bourg2015self}. This directional bias indicates that the simulated uniaxially compacted fabric is more anisotropic than the experimental materials or that the simplified transport mapping amplifies the effect of platelet alignment. Importantly, no transport parameters were fitted to these data: the diffusion trends emerge from the influence of the PMFs on hydration-state populations, pore connectivity and tortuosity.

The mechanical response provides a complementary test of the simulated microstructures. The calculated moduli increase with compaction as non-interlayer pores close and more platelet pairs occupy tightly bound hydration states. The idealized isotropic systems are generally too stiff, whereas the polydisperse assemblies provide softer moduli and better reproduce the experimental modulus--void-ratio trends. The transport and mechanical results therefore favour different aspects of the two simulated fabrics: the isotropic systems provide a useful direction-averaged description of diffusion, while the more strongly oriented polydisperse systems better represent the measured stiffness. This distinction is consistent with the orientation
metrics: the isotropic 12-nm systems remain weakly ordered
(\(S\simeq0.15\)--\(0.18\)), whereas the \(z\)-compressed polydisperse systems
show strong alignment of platelet normals with the compression axis
(\(S_z\simeq0.94\)--\(0.98\)); the full values are reported in the Supplementary
Material. Direct experimental measurements of platelet orientation would be required to determine which simulated fabric most closely represents the tested compacted materials.

Taken together, these results support the central multiscale premise of the study. Accurately representing hydration-state energetics is sufficient to produce compaction-dependent interlayer populations and pore networks, from which transport and elastic trends emerge without separate fitting of those properties. At the same time, the contrasting behaviour of the two platelet ensembles shows that accurate pair energetics alone do not uniquely determine the mesoscale response: platelet-size distribution, preparation history and fabric anisotropy remain important state variables.

Several limitations define the scope of the present model. First, all interactions are represented by distance-dependent two-body terms. The model therefore cannot fully capture orientation-dependent site matching, differences between corners and straight edges or many-body hydration correlations, which likely contribute to the residual errors in rotated and size-mismatched PMFs. Second, the platelets are rigid, excluding bending, edge kinking, delamination and other deformation modes that may alter pore connectivity and mechanical compliance, particularly at high density. Third, water and ions are implicit and the transport analysis considers a neutral tracer. The framework therefore does not describe charged-species partitioning, electrostatic exclusion, diffuse-double-layer effects or changes in electrolyte composition and ion exchange. Finally, the voxelized pore representation may under-resolve narrow throats and rough interfacial voids, contributing to the underestimation of accessible porosity at high compaction.

Within these limitations, the Morse+GPR potential provides a transferable mesoscale description across the Na--MMT configurations and compacted microstructures examined here. It captures the discrete hydration energetics needed to reproduce interlayer-spacing statistics, compaction-driven pore evolution, neutral-tracer diffusion trends and the growth of stiffness with density. The tabulated formulation also provides a practical basis for further development. Priority extensions include introducing edge and orientation descriptors or targeted three-body corrections, allowing platelet flexibility and a broader range of compaction histories, and parameterizing multicomponent systems containing ions such as Na$^+$, Ca$^{2+}$ and Cu$^{2+}$. Coupling these extensions to charged-tracer transport would enable the same multiscale framework to address ion-specific diffusion, partitioning and exchange in compacted swelling clays.


\section*{Acknowledgements}
This work is financially supported by the Nuclear Waste Management Organization, Canada, the Natural Sciences and Engineering Research Council of Canada (NSERC), and Mitacs Canada. The authors also thank the Digital Research Alliance of Canada and the Centre for Advanced Computing at Queen's University for generous allocation of computer resources.

\section*{Data and model availability}
The tabulated Morse+GPR pair potentials used in this work are available in the GitHub repository \url{https://github.com/yaldapedram/nammt-morse-gpr-cg}. Example scripts for PMF processing and model training are also provided. Additional simulation data are available from the corresponding author upon
reasonable request.

\section*{CRediT authorship contribution statement}

Yalda Pedram: Conceptualization, Methodology, Validation, Formal analysis, Investigation, Data curation, Visualization, Writing -- original draft. Yaoting Zhang: Conceptualization, Formal analysis, Validation, Methodology. Laurent Brochard: Formal analysis, Methodology, Validation, Writing -- review \& editing. Chang Seok Kim: Validation, Writing -- review \& editing. Laurent Karim Béland: Conceptualization, Formal analysis, Funding acquisition, Project administration, Supervision, Validation, Writing -- review \& editing.

\section*{Declaration of competing interest}

The authors declare that they have no known competing financial interests or personal relationships that could have appeared to influence the work reported in this paper.

\bibliographystyle{elsarticle-harv}
\bibliography{refs}                 
\end{document}